%% file: synch_2.tex
\documentclass[conference,comsoc]{IEEEtran}

\usepackage{tikz}
\usepackage{float}
\usepackage{amsmath}
\usepackage{amssymb}
\usepackage{commath}
\usepackage{algorithm}
\usepackage{algorithmic}
\usetikzlibrary{bayesnet}
\usepackage{multicol}
\usepackage[noadjust]{cite}
\usepackage{graphicx}
\usepackage{ragged2e}
\usepackage{hyperref}
\usetikzlibrary{decorations.pathmorphing}
\usepackage{caption}
\usepackage{subcaption}
\hypersetup{
	pdfstartview=FitH,
	colorlinks=false,
	linkbordercolor=black,
	pdfborderstyle={/S/U/W 0}
}
\usetikzlibrary{patterns,positioning,fit,shapes,arrows}

\input{Macros}

\begin{document}
\IEEEoverridecommandlockouts
\title{A Hybrid Bayesian Approach Towards Clock Offset and Skew Estimation in 5G Networks
\thanks{The research leading to these results was funded by European Union's Framework Programme Horizon 2020 for research, technological development and demonstration under grant agreement No. 762057 (5G-PICTURE).}}

\author{\IEEEauthorblockN{Meysam Goodarzi\rlap{\textsuperscript{\IEEEauthorrefmark{1}}},\,\textsuperscript{\IEEEauthorrefmark{2}} Darko Cvetkovski\rlap{\textsuperscript{\IEEEauthorrefmark{1}}},\,\textsuperscript{\IEEEauthorrefmark{2}} Nebojsa Maletic\rlap{\textsuperscript{\IEEEauthorrefmark{1}}},\, Jes{\'u}s Guti{\'e}rrez\rlap{\textsuperscript{\IEEEauthorrefmark{1}}}, and Eckhard Grass\rlap{\textsuperscript{\IEEEauthorrefmark{1}}},\,\textsuperscript{\IEEEauthorrefmark{2}}}
\IEEEauthorblockA{\IEEEauthorrefmark{1}IHP -- Leibniz-Institut f\"{u}r innovative Mikroelektronik, Frankfurt (Oder), Germany}
\IEEEauthorblockA{\IEEEauthorrefmark{2}Humboldt University of Berlin, Berlin, Germany.}
Emails:\{goodarzi, cvetkovski, maletic, teran, grass\}$@$ ihp-microelectronics.com}
\maketitle

\begin{abstract}
In this work, we propose a hybrid Bayesian approach towards clock offset and skew estimation, thereby synchronizing large scale networks. In particular, we demonstrate the advantage of Bayesian Recursive Filtering (BRF) in alleviating time-stamping errors for pairwise synchronization. Moreover,  we indicate the benefit of Factor Graph (FG), along with Belief Propagation (BP) algorithm in achieving high precision end-to-end network synchronization. Finally, we reveal the merit of hybrid synchronization, where a large-scale network is divided into local synchronization domains, for each of which a suitable synchronization algorithm (BP- or BRF-based) is utilized. The simulation results show that, despite the simplifications in the hybrid approach, the Root Mean Square Errors (RMSEs) of clock offset and skew estimation remain below 5 ns and 0.3 ppm, respectively.

\end{abstract}
%
\IEEEpeerreviewmaketitle
\section{Introduction}\label{sec:intro}
A large variety of sync\footnote{The words ``synchronization'' and ``sync'' are used alternatively in this paper and carry the same meaning.}-based services such as distributed beamforming \cite{jagannathan2004effect}, tracking \cite{wu2010clock}, mobility prediction \cite{goodarzi2019next}, and localization \cite{etzlinger2017cooperative,koivisto2017joint,zheng2009joint} are expected to be delivered by the fifth generation (5G) of wireless networks. To prepare the fertile ground for these services, considerable effort has been put into designing algorithms for fast, continuous, and precise synchronization \cite{levesque2016survey}. In general, state-of-the-art algorithms achieve synchronization in a network by adopting two macroscopic approaches: a) structurally employing the existing \textit{pairwise synchronization} protocols, e.g. layer-by-layer pairwise synchronization \cite{giorgi2011performance, leng2011low, lv2014simulation},
and b) design an algorithm from scratch to perform \textit{network-wide synchronization} \cite{leng2011distributed, zou2015network, etzlinger2014cooperative, du2013distributed}. 

For pairwise synchronization, IEEE 1588 (often denoted as Precision Time Protocol (PTP) \cite{eidson2002ieee}), is perhaps the most common protocol, deployed in numerous applications. Along with the Best Master Clock Algorithm (BMCA), the PTP utilizes hardware time-stamping and pairwise communication between nodes to determine the Master Node (MN) and, consequently, to perform synchronization in tree-structured networks. While this combination succeeds in networks with medium time precision sensitivity (e.g. sub-$\mu$s range), uncertainty in time-stamping \cite{giorgi2011performance} and BMCA failure in determining the MN \cite{gaderer2008master} can lead to a considerable deterioration of the performance in time precision sensitive networks. The former is caused by the layer where the time-stamps are taken, while the latter can result from the mesh topology of the communication network.
It has been attempted in \cite{giorgi2011performance} to address the time-stamping error by the virtue of Kalman filtering. However, since all the information available by time-stamps is not exploited, the approach is not optimal in the Bayesian sense. Instead, the Bayesian Recursive Filtering (BRF) used in \cite{rhee2009clock} can be employed to capture all the available information in time-stamps, thereby optimally rectifying the time-stamping error. Furthermore, in \cite{leng2011distributed} network-wide synchronization in wireless sensor networks is performed with the help of the Belief Propagation (BP) algorithm running on Factor Graphs (FGs). In BP, in contrast to BMCA, the nodes exchange their information about each other, thereby reaching an agreement about their clock status even if the network (or its corresponding FG) contains loops. Nevertheless, the time required by BP for synchronization is considered to be a potential drawback. 

Despite the valuable contribution made towards synchronization by the aforementioned works, it appears to be unlikely that each individual algorithm can alone achieve the global and local time precision aimed by 5G \cite{ruffini2017novel}. Instead, owing to diverse topology (e.g. tree and mesh) of a network, it is anticipated that a combination of these algorithms would deliver a superior performance when compared to each alone \cite{goodarzi2020synchronization}. In particular, to satisfy the requirements on both the absolute and relative time error in a diversely-structured large-scale network, the architecture of the 5G synchronization network has been suggested to consist of common synchronization areas and various synchronization domains \cite{li2017analysis}. Therefore, a promising approach appears to be equipping the network with different sync algorithms (or a combination thereof), whereby  each domain can leverage a suitable sync algorithm based on its topology and capabilities. In this manner, while keeping the absolute time error low, it is easier to satisfy the requirement of the relative time error in the sync domains.  
%


The contribution of this paper is summarized as follows:
\begin{itemize}
\item We present the principles of pairwise synchronization based on BRF.
\item We develop a network-wide statistical synchronization algorithm based on FG and BP.
\item We adopt a hybrid approach to accurately estimate clock offset and skew, whose performance is then studied by comparing with a non-hybrid algorithm, i.e. BP 
\end{itemize} 
The rest of this paper is structured as follows: In Section II, we introduce our system model. Section III deals with the estimation methods for pairwise, network-wide, and hybrid synchronization. Simulation results are presented and discussed in Section IV. Finally, Section V concludes this work and indicates the future work.
\subsubsection*{Notation} The boldface capital $\boldsymbol{A}$ and lower case $\boldsymbol{a}$ letters denote matrices and vectors, respectively. $\boldsymbol{A}^T$ indicate the transposed of matrix $\boldsymbol{A}$. $\boldsymbol{I}_N$ represents a $N$ dimensional identity matrix. $\mathcal{N}(\mathbf{x}|\boldsymbol{\mu}, \boldsymbol{\Sigma})$ denotes a random vector $\mathbf{x}$ distributed as Gaussian with mean vector $\boldsymbol{\mu}$ and covariance matrix $\boldsymbol{\Sigma}.$ The symbol $\thicksim$ stands for ``is distributed as" and the symbol $\propto$ represents
the linear scalar relationship between two functions. 
\section{System Model}
\subsection{Clock Model}
Each node $i$ is considered to have the clock model 
\begin{equation}
c_i(t) = \gamma_i t + \theta_i,
\label{eq:clkmod}
\end{equation}
where $t$ represents the reference time. Furthermore, $\gamma_i$ and $\theta_i$ denote the clock skew and offset, respectively. In fact, (\ref{eq:clkmod}) determines how the reference time is mapped onto clock of node $i$. The parameter $\gamma_i$ is generally random and varies over time. However, it is common to assume that it stays constant in the course of one sync period \cite{etzlinger2014cooperative,giorgi2011performance}. Moreover, $\theta_i$ is due to several components, all are extensively discussed in the following subsection.  Given that, the goal of time synchronization is to estimate $\gamma_i$ and $\theta_i$ (or transformations thereof) for each node and apply correction such that, ideally, all the clocks show the same time as the reference time $t$. 
\subsection{Offset Decomposition and Measurement Model}\label{ssec:offdec}
\begin{figure}[t!]
\includegraphics[width=0.95\linewidth]{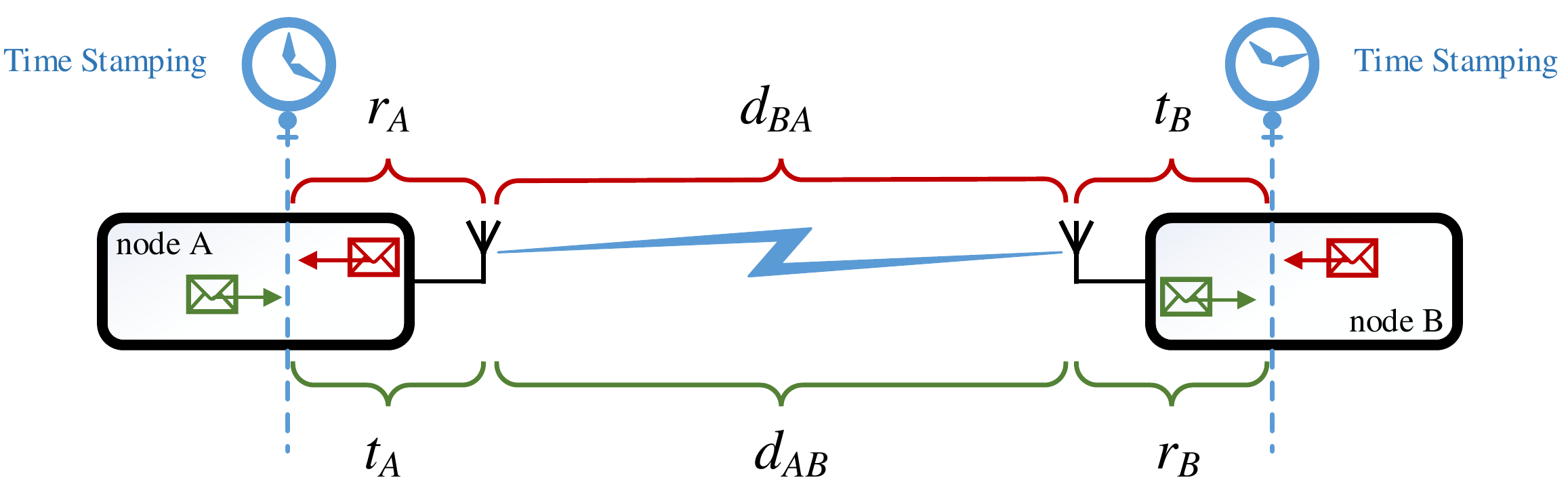}
\centering
\caption{Delay Decomposition.}
\label{fig:deldec}
\end{figure}

To acquire a sensible conception of the components making up the offset $\theta_i$, we decompose it as shown in Figure \ref{fig:deldec}. The parameter $t_A$ (and $t_B$) is the time taken for a packet to leave the transmitter after being time-stamped (the term ``time-stamp'' is refered to hardware time-stamping hereafter), $d_{AB}$ and $d_{BA}$ denote the propagation delay, and $r_B$ (and $r_A$) represents the time that a packet needs to reach the time-stamping point upon arrival at the receiver. In general, the packets sent from node A to node B do not experience the same delay as the packets sent from node B to node A. In other words $$t_A + d_{AB} + r_B \neq t_B + d_{BA} + r_A.$$ Furthermore, we define $T_{AB} = t_A + r_B,$ and $R_{AB} = t_B + r_A$. Generally, $T_{AB}$ and $R_{AB}$ (and correspondingly $t_A,$ $t_B,$ $r_A,$ and $r_B$) are random variables due to several hardware-related random independent processes and can, therefore, be assumed i.i.d. Gaussian random variables, whereas $d_{AB}$ and $d_{BA}$ are usually assumed to be deterministic and symmetric ($d_{AB} = d_{BA}$) \cite{leng2011distributed}. We use the time-stamping mechanism shown in Figure \ref{fig:stamp}, implemented by the PTP protocol \cite{eidson2002ieee}. Thus
\begin{figure}[t!]
\begin{tikzpicture}[scale=1]
	\definecolor{mc}{rgb}{0.4, 0.55, 0.9};
	\definecolor{mc1}{rgb}{0.66, 0.11, 0.03};
	\definecolor{mc2}{rgb}{0.0, 0.44, 0.0};
	
	\draw (-3.5, 0.5) node(cj0){};
	\draw (-3.5, -0.3) node(ci0){};
	\draw (1, 1.) node(cjend){};
	\draw (1, -0.7) node(ciend){};
	\draw[thick,-] (cj0.south)--(cjend.south);
    \draw[thick,-] (ci0.south)--(ciend.south);
    
	\draw (-3.6, 0) node(tbeg){};
	\draw (1.7, 0) node(tend){$c(t)=t$};
	\draw[mc1,thick,dashed] (tbeg.east)--(tend.west);
	
    \draw (-2.8, 0.8) node(cj2){$c_i(t_2^{k})$};
    \draw (-1.9, 0.9) node(cj3) {$c_i(t_3^{k})$};
    \draw (0.5, 1.15) node(cjj){$c_i(t_2^{k+1})$};
    
    \draw (-3.4, -0.8) node(ci1){$c_j(t_1^{k})$ };
    \draw (-1.3, -0.95) node(ci4) {$c_j(t_4^{k})$};
    \draw (0, -1.1) node(cii){$c_j(t_1^{k+1})$};
    
    \draw[mc, thick,->] (ci1.north)--(cj2.south);
    \draw[thick, mc,->] (cj3.south)--(ci4.north);
    \draw[thick, mc,->] (cii.north)--(cjj.south);
    
    \draw (-3.4,-1.5) node(b1){};
    \draw (-2.8,-1.5) node(b2){};
    \draw (-1.9,-1.5) node(e1){};
    \draw (-1.3,-1.5) node(e2){};
    \draw (-3.1,-1.9) node(b3){$d_{ij}+T_{ij}$};
    \draw (-1.6,-1.9) node(b4){$d_{ij}+R_{ij}$};
    \draw[mc2,thick,dashed] (cj2.south)--(b2.north);
    \draw[mc2,thick,dashed] (ci1.north)--(b1.north);
    \draw[mc2,thick,dashed] (ci4.north)--(e2.north);
    \draw[mc2,thick,dashed] (cj3.south)--(e1.north);
\end{tikzpicture}
\centering
\caption{Time-stamp exchange between nodes $i$ and $j$.}
\label{fig:stamp}
\end{figure}
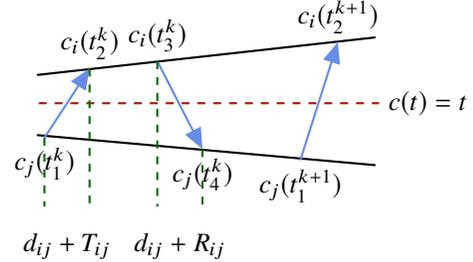
\begin{align}
&\frac{1}{\gamma_i}(c_i(t_{2}^k) - \theta_i) = \frac{1}{\gamma_j} (c_j(t_1^k) - \theta_j) + d_{ij} + \tij{k}  ,\label{eq:c1}\\ 
&\frac{1}{\gamma_i}(c_i(t_{3}^k) - \theta_i) = \frac{1}{\gamma_j} (c_j(t_4^k) - \theta_j) - d_{ij} - \rij{k}. \label{eq:c2}
\end{align}
where $t_1^k$/$t_4^k$ and $t_3^k$/$t_2^k$ are the time points where neighboring nodes $j$ and $i$ send/receive the sync messages, respectively.
By the end of the $k$-th round of time-stamp exchange, each node is expected to have collected the time-stamps  
$\Cij = \begin{bmatrix}
\ccij^1, \cdots, \ccij^k
\end{bmatrix}^{T},$ where $$\ccij^k = \left[c_j(t_1^k),  c_i(t_2^k), c_i(t_3^k), c_j(t_4^k)\right]. $$
\section{Clock Offset and Skew Estimation}
In this section, we firstly introduce BRF-based pairwise synchronization. Subsequently, we describe the principles of network-wide synchronization based on BP. Lastly, we present an approach, where both techniques are employed in a hybrid manner.
\subsection{Pairwise Offset and Skew Estimation}\label{ssec:brf}
\begin{figure}[t!]
\begin{tikzpicture}
    \draw (0, 0) node[inner sep=0] {\includegraphics[width=0.85\linewidth]{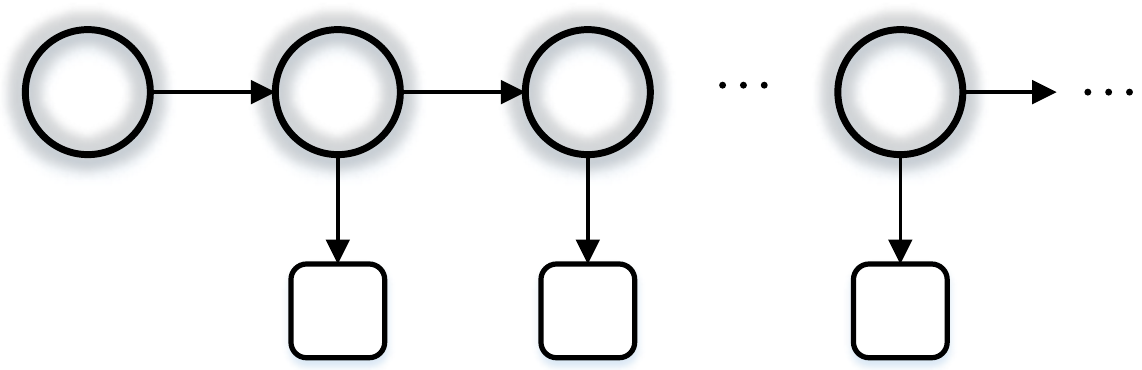}};
	\draw (-3.15,0.6) node(a00){$\tetvect{i}^0$};  
	\draw (-1.5,0.6) node(a00){$\tetvect{i}^1$};
	\draw (.15,0.6) node(a00){$\tetvect{i}^2$};
	\draw (2.2,0.6) node(a00){$\tetvect{i}^k$};
	
	\draw (-1.5,-0.85) node(a00){$\ccij^1$};
	\draw (.15,-0.85) node(a00){$\ccij^2$};
	\draw (2.20,-0.85) node(a00){$\ccij^k$};
\end{tikzpicture}
\centering
\caption{Bayesian representation of offset and skew estimation.}
\label{fig:bayesrep}
\end{figure}
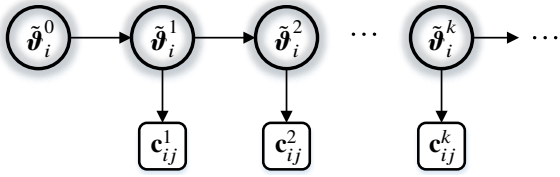

In pairwise synchronization, one node is assumed to be the MN\footnote{In Figure \ref{fig:stamp}, instead of a global reference $c(t)=t,$ we take node $j$ as MN. It is straightforward to see that $\frac{1}{\tilde{\gamma}_i}=\frac{\gamma_j}{\gamma_i},$ $\tilde{\theta}_i = \theta_i-\tg\theta_j,$ $\tilde{d}_{ij} + \tijt{k} = \gamma_j(d_{ij} + \tij{k}),$ and $\tilde{d}_{ij} - \rijt{k} = \gamma_j(d_{ij} - \rij{k})$. For the sake of simplicity, as done in \cite{wu2010clock}, we assume $\tilde{d}_{ij}=d_{ij},$ $\rijt{k} = \rij{k},$ and $\tijt{k}=\tij{k}.$ This is valid owing to $\gamma_j\approx 1$ and the value of $d_{ij} + \tij{k}$ and $d_{ij} - \rij{k}$ being low.}. Consequently (\ref{eq:c1}) and (\ref{eq:c2}) turn into
\begin{align}
&\tgam(c_i(t_{2}^k) - \tilde{\theta}_i) = c_j(t_1^k) + d_{ij} + \tij{k}  ,\label{eq:c11}\\ 
&\tgam(c_i(t_{3}^k) - \tilde{\theta}_i) = c_j(t_4^k) - d_{ij} - \rij{k}. \label{eq:c22}
\end{align}
Let $\ttat{k}$ be the state of the vector variable $\tetvect{i} \triangleq \left[\tgam, \frac{\ttet}{\tg}\right]^T$ after $k$-th round of time-stamp exchange (visualized in Figure \ref{fig:bayesrep}). 
The probability distribution function (pdf) corresponding to $k$-th state can then be written as
\begin{equation}
p(\ttat{k}|\Cij) = \int p(\ttat{0},\cdots, \ttat{k}|\Cij)\ d\Theta^{k-1},
\end{equation}
where $\Theta^{k-1} = \left[\ttat{0},\cdots,\ttat{k-1}\right]$. Employing Bayes rule: 
\begin{equation}
p(\ttat{k}|\Cij) \propto \int p(\Cij|\ttat{0},\cdots, \ttat{k})p(\ttat{0},\cdots, \ttat{k})\ d\Theta^{k-1}.
\label{eq:bayesrule}
\end{equation}
Assuming the independent measurements and Markov property \cite{barker1995bayesian}, the integrands in (\ref{eq:bayesrule}) can be rewritten as 
\begin{align}
\begin{split}
&p(\Cij|\ttat{0},\cdots, \ttat{k}) = \cntnt{k}{k}\cdots \cntnt{1}{1}, \\
&p(\ttat{0},\cdots, \ttat{k}) = \tntt{k}{k-1}\cdots \tntt{1}{0}p(\tetvect{i}^0),
\end{split}
\label{eq:markov}
\end{align}
where $p(\tetvect{i}^0)$ denotes the prior knowledge on $\tetvect{i}.$ Plugging (\ref{eq:markov}) into (\ref{eq:bayesrule}) leads to
\begin{multline}
p(\ttat{k}|\Cij) \propto  \\ \small \underbrace{\int p(\ttat{0})\left[\prod_{r=1}^{k-1}p(\ttat{r}|\ttat{r-1})\cntnt{r}{r}\right]\tnt{k}{k-1}d\Theta^{k-1}}_{= p(\tta{k}|\ccij^{1:k-1})}\cntnt{k}{k}, \\
\label{eq:longeq}
\end{multline}
which can then be simplified as follows:
\begin{align}
p(\ttat{k}|\Cij) \propto p(\tetvect{i}^{k}|\ccij^{1:k-1})\cntnt{k}{k} \sim \mathcal{N}(\boldsymbol{\mu}_i^k, \mathbf{Q}_i^k).
\label{eq:bayesfin}
\end{align}
The term $p(\tetvect{i}^{k}|\ccij^{1:k-1})$ is referred to as \textit{prediction} step while the term $\cntnt{k}{k}$ is considered as \textit{measurement update} step \cite{barker1995bayesian}. In wireless networks, due to clock properties, it is typical to assume that $\tetvect{i}^k$ is Gaussian distributed \cite{wu2010clock,etzlinger2014cooperative,etzlinger2017cooperative}. Given this assumption, in the sequel, we show that the relation between the states is linear, and therefore, the marginal in (\ref{eq:bayesfin}) is also Gaussian distributed.
\subsubsection{Prediction} Assuming constant skew in one synchronization period ($=$ $K$ rounds of time-stamp exchange), a reasonable prediction for $\tetvect{i}^k$ is given by \cite{giorgi2011performance},  

\begin{equation}
\tetvect{i}^k = \mathbf{A}\tetvect{i}^{k-1}+\mathbf{u}^{k-1}_i+\mathbf{n}^{k-1}_i,
\label{eq:predlin}
\end{equation}
where $\mathbf{A}=\begin{bmatrix} 1 & 0 \\  c_j(t_{1}^{k})-c_j(t_{1}^{k-1})  & 1 \end{bmatrix},$
and $\mathbf{u}^{k-1}_i = \left[0, -\frac{1}{\tg^{k-1}}\left(c_j(t_{1}^{k})-c_j(t_{1}^{k-1}) \right)\right]^T$ is input correction vector and removes the impact of time evolution when predicting $\frac{\ttet^k}{\tg^k}$. Moreover, $\mathbf{n}^{k-1}_i$ denotes the Gaussian noise vector and is assumed to be negligible.
Given (\ref{eq:predlin}), the prediction term can be written as
\begin{equation}
 p(\tetvect{i}^{k}|\ccij^{1:k-1}) \sim \mathcal{N}(\tetvect{i}^k | \mupred, \copred),
 \label{eq:predpdf}
\end{equation}
where $\mupred  = \mathbf{A}\boldsymbol{\mu}_i^{k-1} + \mathbf{u}^{k-1}_i$ and $\copred = \mathbf{A}\mathbf{Q}^{k-1}_i\mathbf{A}^T.$
\subsubsection{Measurement update} We conduct the following mathematical manipulations to obtain the update term in (\ref{eq:bayesfin}). Subtracting (\ref{eq:c11}) in the $(k-1)$-th round from that of the $k$-th round leads to
\begin{align}
& \frac{1}{\tilde{\gamma}_i}(c_i(t_{2}^{k}) - c_i(t_{2}^{k-1})) = c_j(t_1^{k}) - c_j(t_1^{k-1}) + \tij{k}-\tij{k-1}, \label{eq:c1-c2}
\end{align} 
while summing up (\ref{eq:c11}) and (\ref{eq:c22}) in the $k$-th round gives 
\begin{align}
&\tgam(c_i(t_{2}^k) + c_i(t_{3}^k)-2\ttet) = c_j(t_1^k) + c_j(t_4^k) + \tij{k}-\rij{k}, \label{eq:c1+c2} 
\end{align}
where $\tij{k} - \rij{k}$ and  $\tij{k}-\tij{k-1}$ are assumed to be zero mean and have the variance $\sigma^2_{T_{ij}} + \sigma^2_{R_{ij}}$ and $2\sigma^2_{T_{ij}},$ respectively. This is straightforward to observe since they are linear subtraction of independent random processes. The parameters $\sigma^2_{T_{ij}}$ and $\sigma^2_{R_{ij}}$ are mostly related to the hardware properties of the nodes and assumed to be static and known \cite{leng2011distributed,etzlinger2014cooperative}.  
Alternatively, we can write (\ref{eq:c1-c2}) and (\ref{eq:c1+c2}) in matrix form as
\begin{align}
\mathbf{B}_{ij}\tetvec{i}  = \mathbf{r}_{ij} + \zji,
\end{align}
where $\zji\sim \mathcal{N}(\mathbf{z}|\mathbf{0},\mathbf{R}_{ij})$ with $\mathbf{R}_{ij} = \begin{bmatrix}
2\sigma^2_{T_{ij}} & 0 \\ 0 & \sigma^2_{T_{ij}} + \sigma^2_{R_{ij}} \end{bmatrix},$
 $\mathbf{B}_{ij} = \begin{bmatrix}
 c_i(t_{2}^{k}) - c_i(t_{2}^{k-1}) & 0 \\
 c_i(t_{2}^k) + c_i(t_{3}^k) & -2
\end{bmatrix},$ \\and $\mathbf{r}_{ij} = \left[
c_j(t_1^{k}) - c_j(t_1^{k-1}), c_j(t_{1}^k) + c_j(t_{4}^k) 
\right]^T.$
\\
Consequently,
\begin{equation}
\cntn{k}{k} \sim \mathcal{N}(\mucorr, \cocorr),
\label{eq:mespdf}
\end{equation}
where $\mucorr = \mathbf{B}_{ij}^{-1}\mathbf{r}_{ij}$ and $\cocorr = \mathbf{B}_{ij}^{-1}\mathbf{R}_{ij}\mathbf{B}_{ij}^{-T}$.
\subsubsection{Estimation} Considering (\ref{eq:predpdf}) and (\ref{eq:mespdf}), the estimated distribution in (\ref{eq:bayesfin}) is given by
\begin{equation}
p(\ttat{k}|\Cij)\sim \mathcal{N}(\muest, \coest),
\label{eq:estpdf}
\end{equation}
where
\begin{align}
&\muest = \left[\copred + \cocorr\right]^{-1}\left(\cocorr\mupred + \copred\mucorr\right), \\
&\coest = \left[\copred^{-1} + \cocorr^{-1}\right]^{-1}.
\end{align}
The parameters in (\ref{eq:predpdf}), (\ref{eq:mespdf}), and (\ref{eq:estpdf}) are calculated recursively and, in each iteration $k,$ the estimation of the clock offset and skew can be obtained by
\begin{equation}
\tg^k = \frac{1}{\muest(1)}\ \text{and}\ \ttet^k = \frac{\muest(2)}{\muest(1)},
\label{eq:finestpair}
\end{equation}
where $\muest(1)$ and $\muest(2)$ are the first and second element of the vector $\muest$, respectively. 
Algorithm \ref{alg:brf} summarizes this recursive process.
\begin{algorithm}[t!]
\begin{algorithmic}[1]
\STATE Initialize p($\tetvect{i}^0$) to be non-informative\label{init}
\FOR {$k = 1, 2, \cdots, K$}  \label{forBRF}
\STATE Calculate the mean vector and covariance matrix of the \textit{prediction} pdf using (\ref{eq:predpdf}) \label{predpdf}
\STATE Construct $\mathbf{B}_{ij},$ $\mathbf{R}_{ij},$ and $\mathbf{r}_{ij}$ using the measurements and obtain the mean vector and covariance matrix of \textit{update} pdf using (\ref{eq:mespdf})\label{mespdf}
\STATE Compute the mean vector and covariance matrix of the pdf of $\tetvect{i}^k$ using (\ref{eq:estpdf})\label{finpdf}
\ENDFOR \label{endfor}
\STATE Compute the final estimation of offset and skew using (\ref{eq:finestpair})
\end{algorithmic}
\caption{Pairwise synchronization based on BRF}
\label{alg:brf}
\end{algorithm}
%
\subsection{Network-wide Offset and Skew Estimation}\label{ssec:fg}
Unlike pairwise sync, in network-wide sync we aim to synchronize each node with a global MN. Therefore, the statistical model obtained in \ref{ssec:brf} based on relative offset and skew is insufficient for network-wide synchronization. In the sequel, we obtain the pairwise statistical model assuming that both clocks have offset and drift relative to a global MN.
\subsubsection{Pairwise statistical model}
Summing up (\ref{eq:c1}) and (\ref{eq:c2}) and stacking the resulting equations for $K$ rounds of time-stamp exchange, we can write
\begin{equation}
\aji\tetvec{i} + \aij\tetvec{j} = \zji,
\label{eq:veceq}
\end{equation}
where $\aji$ and $\aij$ are $K\times 2$ matrices with the $k$-th row being $\left[c_i(t_{2}^k)+c_i(t_{3}^k), -2\right]$ and $-\left[c_j(t_{1}^k)+ c_j(t_{4}^k), -2\right]$, respectively. Moreover, similar to \ref{ssec:brf}, we introduce the vector variables $\tetvec{i}\triangleq\left[\frac{1}{\gamma_i}, \frac{\theta_i}{\gamma_i}\right]^T,$ and $\tetvec{j}\triangleq\left[\frac{1}{\gamma_j}, \frac{\theta_j}{\gamma_j}\right]^T$ with
$\frac{1}{\gamma_i},$ $\frac{\theta_i}{\gamma_i},$ $\frac{1}{\gamma_j},$ and $\frac{\theta_j}{\gamma_j}$ being Gaussian distributed \cite{etzlinger2017cooperative}.
Finally $\zji\sim \mathcal{N}(\mathbf{z}|\mathbf{0}, \sigma_{ij}^2\mathbf{I}_K),$ where $\sigma_{ij}^2 = \sigma_{T_{ij}}^2 + \sigma_{R_{ij}}^2.$
What (\ref{eq:veceq}) implicitly states is that for given $\tetvec{i}$ and $\tetvec{j},$ the probability that we measure $\aji$ and $\aij$ is equal to $\mathcal{N}(\mathbf{z}=\aji\tetvec{i} + \aij\tetvec{j}|\mathbf{0}, \sigma_{ij}^2\mathbf{I}_N),$ what can be expressed as
\begin{equation}
p(\aji, \aij|\tetvec{i}, \tetvec{j}) \sim \mathcal{N}(\mathbf{z}=\aji\tetvec{i} + \aij\tetvec{j}|\mathbf{0}, \sigma_{ij}^2\mathbf{I}_N).
\label{eq:likfun}
\end{equation} 
The aim is to estimate $\gamma_i$ and $\theta_i$ or, alternatively, $\tetvec{i}$, based on the observation matrices $\aji$ and $\aij$. To this end, we rely on Bayesian estimation given by
\begin{multline}
p(\tetvec{i}|\aji, \aij) = \int p(\tetvec{i}, \tetvec{j}|\aji, \aij) d\tetvec{j}\\
\propto \int p(\aji, \aij|\tetvec{i}, \tetvec{j})p(\tetvec{i})p(\tetvec{j})d\tetvec{j},
\label{eq:estpair}
\end{multline}
where $p(\tetvec{i})$ and $p(\tetvec{j})$ denote the Gaussian distributed prior knowledge on $\tetvec{i}$ and $\tetvec{j}$, respectively.
Extending (\ref{eq:estpair}) for the whole network, we obtain the posterior distribution as
\begin{multline}
p(\tetvec{i}|\{\aji,\aij\}_{i=1:M, j\in ne(i)}) =\\ \int\cdots \int p(\tetvec{1}, \cdots, \tetvec{M}|\{\aji,\aij\}_{i=1:M, j\in\ ne(i)})\\ d\tetvec{1}\cdots d\tetvec{i-1}d\tetvec{i+1}\cdots d\tetvec{M},
\label{eq:estnet}
\end{multline}
where $ne(i)$ represents the set of neighboring nodes of node $i$ and $M$ is total number of the nodes in the network. 
Consequently, the estimation of $\tetvec{i}$ can be calculated as
\begin{equation}
\hat{\boldsymbol{\vartheta}}_i = \argmax_{\tetvec{i}}p(\tetvec{i}|\{\aji,\aij\}_{i=1:M, j\in ne(i)}).
\label{eq:est}
\end{equation}
In general, the computation of the marginal pdf in (\ref{eq:estnet}) is costly and of NP-hard complexity. However, the conditional probability under the integral of (\ref{eq:estnet}) can be approximated using \textit{variational} procedure described in the sequel.
\subsubsection{Variational representation}\label{sssec:variational}
Variational methods can approximate an intractable complex distribution $p(x)$ by a simpler straightforward distribution $q(x)$. A popular way to do that is to minimize the discrepancy measure Kullback-Leibler (KL) divergence between $p(x)$ and $q(x)$. It is given by \cite{barberBRML2012}
\begin{equation}
D_{KL}(p\Vert q) = \int_{-\infty}^{+\infty}p(x)\log\left(\frac{p(x)}{q(x)}\right)dx.
\end{equation}
The following structure known as \textit{Bethe free energy} is suggested by statistical physics \cite{zdeborova2016statistical} to be imposed on $q(x)$ in order to minimize KL divergence. That is,
\begin{equation}
q(x) \propto \prod_i q(x_i) \prod_{i,j} q(x_i,x_j),
\end{equation}
with $x_j$ and $x_i$ being neighboring nodes. It turns out that FG can appropriately represent the above structure and BP can efficiently compute the marginal beliefs \cite{barberBRML2012}. Therefore, in the sequel, we introduce FG and BP algorithm.
\subsubsection{Factor Graph}
\begin{figure}[t!]
\includegraphics[width=1\linewidth]{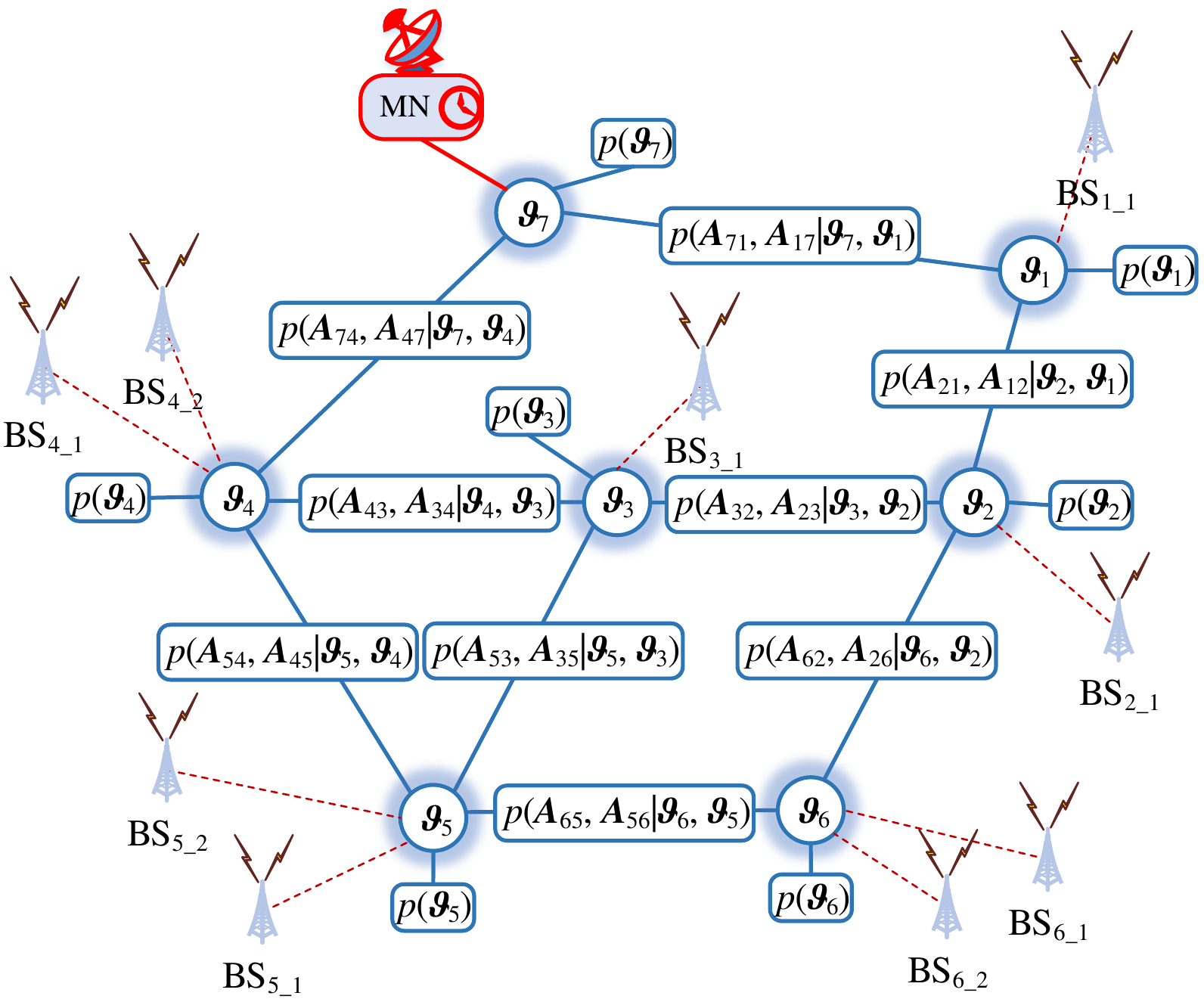}
\centering
\caption{FG corresponding to an exemplary network.}
\label{fig:fg}
\end{figure}

FGs are bipartite graphs used to represent the factorization of a pdf. A FG consists of a number of nodes, each represented by a variable, and several factor nodes, each being a function of its neighboring variables (Figure \ref{fig:fg}). In particular, the factorization and graph structure in FGs can alleviate the computation load, e.g. that of marginal distribution through sum-product algorithm \cite{kschischang2001factor}.
Employing FG and drawing on the idea in \cite{leng2011distributed}, we construct the graphical model in Figure \ref{fig:fg}, where a number of Base Stations (BSs) are backhauled by a mesh network, each node of which is represented by $\tetvec{i}$. 
The goal is then to calculate the marginal of $\tetvec{i}$ using (\ref{eq:estnet}). 

Based on the method outlined in \ref{sssec:variational}, we can approximate the conditional probability under the integral of (\ref{eq:estnet}) as
\begin{multline}
p(\tetvec{1}, \cdots, \tetvec{M}|\{\aji,\aij\}_{i=1:M, j\in\ ne(i)})\propto  \\  \prod p(\tetvec{i}) \prod p(\aij,\aji|\tetvec{i}, \tetvec{j}),
\label{eq:geq}
\end{multline}
where $p(\aji,\aij|\tetvec{i}, \tetvec{j})$ is obtained using (\ref{eq:likfun}).
In the sequel, we briefly illustrate the principles of BP as an efficient algorithm to obtain the estimation in (\ref{eq:est}).
\begin{figure}[t!]
\begin{tikzpicture}[scale=1]
    \draw (0, 0) node[inner sep=0] {\includegraphics[width=0.9\linewidth]{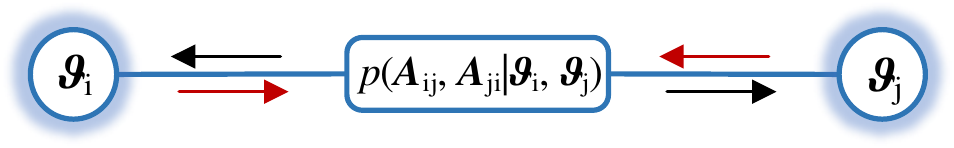}};
	\draw (-1.9,0.45) node(a00){$\delta_{p_{ij}\rightarrow\tetvec{i}}$};  
	\draw (-1.9,-0.45) node(a00){$\delta_{\tetvec{i}\rightarrow p_{ij}}$};
	\draw (2.1,0.45) node(a00){$\delta_{\tetvec{j}\rightarrow p_{ij}}$};
	\draw (2.1,-0.45) node(a00){$\delta_{p_{ij}\rightarrow\tetvec{j}}$};
\end{tikzpicture}
\centering
\caption{Message passing principles in Belief Propagation.}
\label{fig:bpdet}
\end{figure}
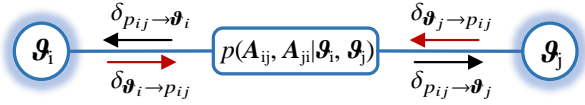
\subsubsection{Belief Propagation}
BP relies on exchanging beliefs between neighboring nodes to compute the marginals. Figure \ref{fig:bpdet} depicts the principles of the message passing in BP algorithm for the nodes $\tetvec{i}$ and $\tetvec{j}$. For the sake of simplicity, we denote the factor $p(\aji, \aij|\tetvec{i}, \tetvec{j})$ with $p_{ij}$. The message from a factor vertex $\pij$ to a variable vertex $\tetvec{i}$ in iteration $l$ is then given by \cite{barberBRML2012}
\begin{multline}
\mij{\pij}{\tetvec{i}}{l}(\tetvec{i}) = \int\ p(\aji, \aij|\tetvec{i}, \tetvec{j}) \mij{\tetvec{j}}{\pij}{l}(\tetvec{j})d\tetvec{j}.
\label{eq:bpmsg1}
\end{multline}
where $\mij{\tetvec{j}}{\pij}{l}(\tetvec{j})$ denotes the message from a variable vertex $\tetvec{j}$ to the variable vertex $\pij$ and is given by
\begin{equation}
\mij{\tetvec{j}}{\pij}{l}(\tetvec{j}) = p(\tetvec{j})\prod_{k\in\{ne(j)\setminus i\}} \mij{p_{kj}}{\tetvec{j}}{l-1}(\tetvec{j})
\label{eq:bpmsg2}
\end{equation}
It is straightforward to see that
\begin{equation}
b^{(l)}(\tetvec{i}) \propto  p(\tetvec{i}) \prod_{p_{ik}\in ne(\tetvec{i})} \mij{p_{ik}}{\tetvec{i}}{l}(\tetvec{i})
\end{equation} 
where $b^{(l)}(\tetvec{i})$ denotes the marginal belief of variable node $\tetvec{i}$ in $l$-th iteration. The outcome of integral in (\ref{eq:bpmsg1}) is expected to be a Gaussian function since its arguments are both Gaussian distributed.
The BP procedure can be summarized as  
\begin{enumerate}
\item The message $\mij{\tetvec{i}}{\pij}{l}(\tetvec{i})$ is transmitted from $\tetvec{i}$ to the neighboring factor nodes $\pij$ (they are initialized non-informatively in the first iteration),
\item The factor node $\pij$ computes the message $\mij{\pij}{\tetvec{i}}{l}(\tetvec{i})$ based on the its incoming messages and sends the calculated messages to the neighboring node $\tetvec{i},$
\item Each node updates its belief based on the received messages from the neighboring factor nodes.
\end{enumerate}
We note that, in practice, there are neither factors nor variable nodes, therefore both (\ref{eq:bpmsg1}) and (\ref{eq:bpmsg2}) are locally computed at each node and only $\mij{p_{ij}}{\tetvec{i}}{l}(\tetvec{i})$ is transmitted from node $j$ to node $i$. Specifically, we can let $\mij{j}{i}{l}(\tetvec{i}) \thicksim \mathcal{N}(\tetvec{i}|\muvec{j}{i}{l}, \sig{j}{i}{l}) $ represent the message sent from $j$ to $i$. Considering (\ref{eq:bpmsg1}) and (\ref{eq:bpmsg2}) together, the covariance matrix $\sig{j}{i}{l}$ can be calculated as \cite{du2013distributed,shental2008gaussian}
\begin{equation}
\sig{j}{i}{l} = \left[\aji^T \left(\omg{j}{i}{l-1}\right)^{-1}\aji\right]^{-1}
\label{eq:var}
\end{equation}
where 
\begin{equation}
\omg{j}{i}{l-1} = \sigma_{ij}^2\uni{N} + \aij \left[\sigsel{j}{-1} + \sum_{k\in ne(j)\setminus i}\left(\sig{k}{j}{l-1}\right)^{-1}\right]^{-1}\aij^T,
\end{equation}
and $\sigsel{j}{}$ is the covariance matrix of $p(\tetvec{j})$. Furthermore,
\begin{multline}
\muvec{j}{i}{l} = -	\sig{j}{i}{l}\aji^T\omg{j}{i}{l-1}\aij \left[\sigsel{j}{-1} + \sum_{k\in ne(j)\setminus i}\left(\sig{k}{j}{l-1}\right)^{-1}\right]^{-1} \\ \times\left[\sigsel{j}{-1}\boldsymbol{\mu}_j + \sum_{k\in ne(j)\setminus i}\left(\sig{k}{j}{l-1}\right)^{-1}\muvec{k}{j}{l-1}\right] ,
\label{eq:mean}
\end{multline}
where $\boldsymbol{\mu}_j$ indicates the mean vector of $p(\tetvec{j}).$
 It is noteworthy that $\sigsel{j}{}$ and $\boldsymbol{\mu}_j$ stay constant and do not change during the message updating process.

The BP algorithm initializes the message
from node $j$ to node $i$ as $\mij{j}{i}{0}(\tetvec{i}) \thicksim \mathcal{N}(\tetvec{i}|\mathbf{0}, +\infty\uni{2})$. Each node $j$ computes its outgoing messages according to
(\ref{eq:var}) and (\ref{eq:mean}) in iteration $l$ with its available $\sig{k}{j}{l-1}$ and $\muvec{k}{j}{l-1}$. The belief of node $i$ is then computed as
\begin{equation}
b^{(l)}(\tetvec{i}) \thicksim \mathcal{N}(\tetvec{i}|\belmu{i}{l}, \belcov{i}{l}),
\label{eq:bel}
\end{equation}
where 
\begin{equation}
\belcov{i}{l} = \left[\sigsel{j}{-1} + \sum_{k\in ne(j)\setminus i}\left(\sig{k}{j}{l-1}\right)^{-1}\right]^{-1},
\label{eq:pbel}
\end{equation}
and
\begin{equation}
\belmu{i}{l} = \belcov{i}{l}\left[\sigsel{j}{-1}\boldsymbol{\mu}_j + \sum_{k\in ne(j)\setminus i}\left(\sig{k}{j}{l-1}\right)^{-1}\muvec{k}{j}{l-1}\right].
\label{eq:nubel}
\end{equation}
Finally, the skew and offset estimation can be computed by 
\begin{align}
\hat{\gamma}_i^{(l)} = \frac{1}{\belmu{i}{l}(1)},&\
\hat{\theta}_i^{(l)} = \frac{\belmu{i}{l}(2)}{\belmu{i}{l}(1)},
\label{eq:finest}
\end{align}
where $\belmu{i}{l}(1)$ and $\belmu{i}{l}(2)$ denote the first and second element of vector $\belmu{i}{l},$ respectively.
\subsection{Hybrid BRF-BP}\label{ssec:hyb}
Given Sections \ref{ssec:brf} and \ref{ssec:fg}, we can run the BRF algorithm at the edge of the network where fast and frequent synchronization is required to keep the relative time error low, what is crucial to a number of applications such as localization. Moreover, for the synchronization of backhaul nodes, BP can be used to ensure that the end-to-end time error requirement is fulfilled. 

Algorithm \ref{alg:hybalg} describes the steps of the hybrid synchronization approach. First, in step \ref{detalg} we decide on the network sections where BP and BRF are to be applied (they are labeled as BP-nodes and BRF-nodes, respectively). Later, in step \ref{klex}, the time-stamp exchange mechanism shown in Figure \ref{fig:stamp} and, correspondingly, the BRF algorithm is initiated at BRF-nodes. In step \ref{msgex}, the time-stamp exchange is initiated among the BP-nodes, thereby obtaining the required time-stamps to form the matrices $\aji$ and $\aij$. The BP iterations begin at step \ref{for} and continue until it converges or the maximum number of iterations $L$ is achieved. In step \ref{calmsg}, each BP-node calculates its outgoing messages using (\ref{eq:var}) and (\ref{eq:mean}) and sends them to its corresponding node. Each node's belief and estimations can then be updated in step \ref{caloff} using (\ref{eq:bel}) and (\ref{eq:finest}), respectively. Steps \ref{conv}-\ref{convend} are responsible to check the convergence by comparing the difference between clock offset and skew estimation in iterations $(l)$ and $(l-1)$ with a predefined small value $\epsilon$. It is noteworthy that step \ref{klex} and steps \ref{msgex}-\ref{endfor} can run simultaneously.
\begin{algorithm}[t!]
\begin{algorithmic}[1]
\STATE Determine the suitable algorithm for each part of the network (BP-nodes or BRF-nodes). \label{detalg}
\STATE Start the time-stamping exchange and initiate the algorithm \ref{alg:brf} at BRF-nodes. \label{klex}
\STATE Start the time-stamp exchange between adjacent BP-nodes and construct $\aji$ and $\aij$ for each pair. \label{msgex}
\FOR {$l = 1, 2, \cdots, L$}  \label{for}
\STATE Compute the messages using (\ref{eq:var}) and (\ref{eq:mean}) for each BP-node and transmit them to its neighboring nodes. \label{calmsg}
\STATE Update the belief at each BP-node using (\ref{eq:bel}) and compute the offset and skew estimation using (\ref{eq:finest}). \label{caloff}
\IF{$\hat{\boldsymbol{\vartheta}}_i^{(l)}-\hat{\boldsymbol{\vartheta}}_i^{(l-1)}\leq\epsilon\ \forall i$} \label{conv} 
\STATE Go to step \ref{msgex}.
\ENDIF \label{convend} 
\ENDFOR \label{endfor}
\end{algorithmic}
\caption{Network synchronization algorithm}
\label{alg:hybalg}
\end{algorithm}
\section{Simulation Results}
In our simulations, the network in Figure \ref{fig:fg} is considered as an exemplary scenario, where a number of BSs are backhauled by a wireless mesh network.  We conduct two sets of simulations: a) synchronization of the whole network based on FG only and, correspondingly, on the BP algorithm (the BSs in Figure \ref{fig:fg} are assumed to be variable nodes and connected to the mesh network via factors), and b) synchronization in a hybrid manner, where the mesh backhauling network is synchronized based on BP while the BSs at the edge of the network are being synchronized using BRF. We then compute the Root Mean Square Error (RMSE) of both clock offset and skew estimations as a measure to evaluate the performance. In fact, scenario (a) is considered as the baseline for comparison with the hybrid approach. For the sake of simplicity and without loss of generality, we consider only the nodes $\tetvec{1}$ and $\tetvec{6}$ and their corresponding BSs. Moreover, the simulation parameters are set as in Table \ref{tab:sim} and $\tetvec{7}$ is set to be the MN.
\begin{table}[t!]
\centering
\caption{Simulation parameters}
\begin{tabular}{l|c}
Number of independent simulations &	10000 \\ \hline
Initial random delays &	[-1000, 1000] ns \\ \hline
Number of time-stamp exchange $K$&	10 \\ \hline
Standard deviation of $T^k_{ij}$ and $R^k_{ij}$ &	4 ns \\ \hline
Random delay between each pair of nodes &	$\left[ 200, 300 \right]$ ns \\ \hline
Initial pdf of the offset/skew for each node & $\mathcal{N}(0,+\infty)$/$\mathcal{N}(1,10^{-4})$\\\hline
Initial pdf of the offset/skew of MN & $\mathcal{N}(0,0)$/$\mathcal{N}(1,0)$
\end{tabular}
\label{tab:sim}
\end{table}

\begin{figure*}[t!]
\begin{subfigure}[]{0.5\textwidth}
\includegraphics[width=.82\linewidth]{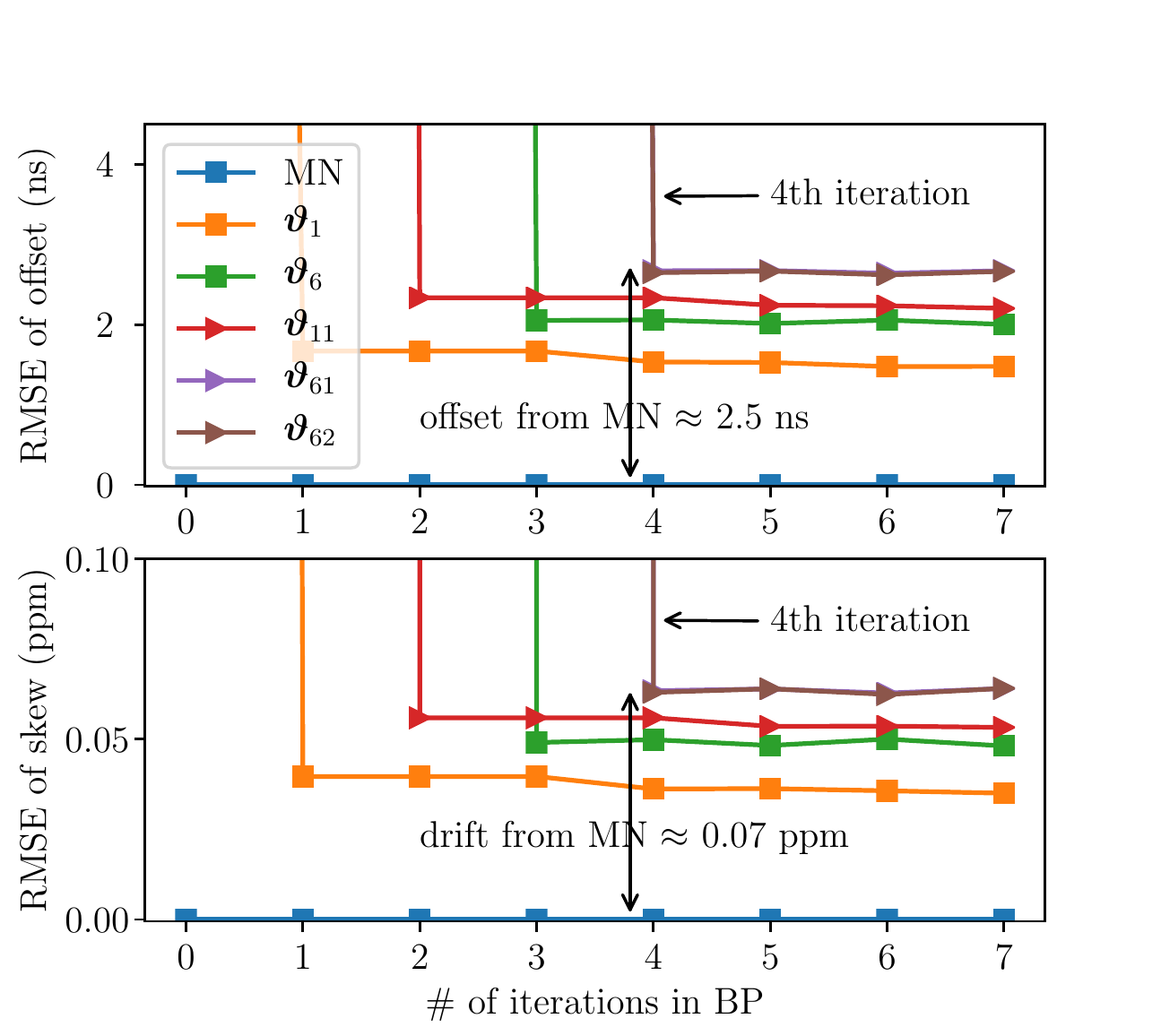}
\centering
\caption{BP applied on the whole network.}
\label{fig:bpwhole}
\end{subfigure}
\begin{subfigure}[]{0.5\textwidth}
\includegraphics[width=0.82\linewidth]{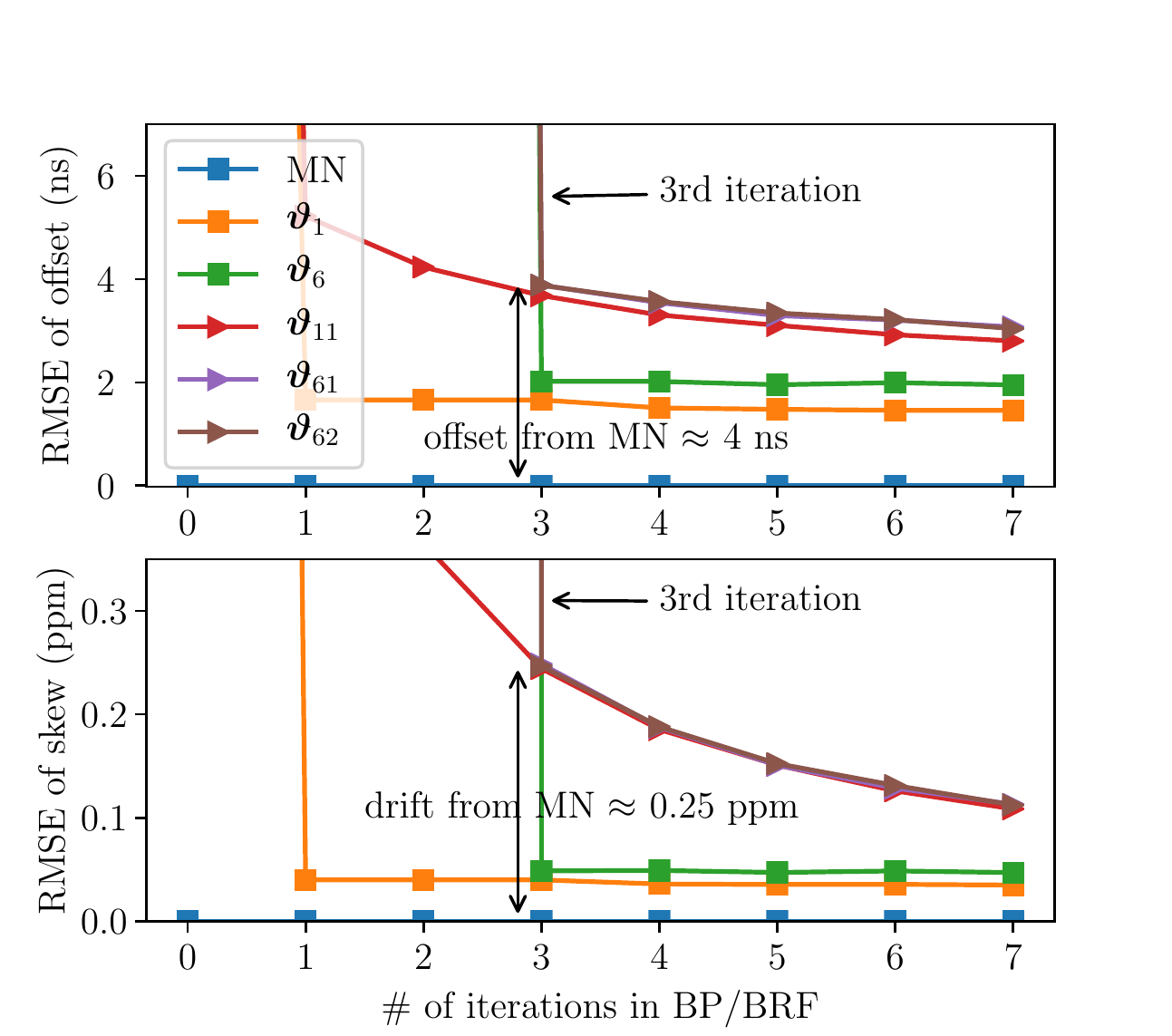}
\centering
\caption{BP and BRF applied to the network in a hybrid manner.}
\label{fig:bpmesh}
\end{subfigure}
	\caption{Performance of sync algorithms applied on the network in Figure \ref{fig:fg}. $0$-th iteration = root mean square of initial values.}
\vspace{0.04cm}
\hrule
\end{figure*}

Figure \ref{fig:bpwhole} represents the RMSE of offset and skew estimation  versus the number of iterations for scenario (a). The RMSEs of offset and skew are represented in \textit{nanosecond} (ns) and \textit{part per million} (ppm), respectively.  As can be seen, the BP converges after $4$ iterations for both offset and skew estimation. The convergence is guaranteed for networks with at least one MN \cite{leng2011distributed}. However, when a network contains loops, the value to which BP converges, is considered to be approximate \cite{barberBRML2012}. Besides, BP achieves an offset RMSE below $3$ ns while that of skew is kept below $0.1$ ppm. 
In fact, the results in this simulation setup reveal the potential performance of BP for time synchronization in communication networks. However, the nodes, and particularly the BSs, must wait at least $4$ iterations (in addition to $K$ time-stamp exchange rounds required for the nodes to obtain the statistics) to be completely synchronized. This can be troublesome in certain sync-based services, e.g. localization, where continuous time alignment is essential. Therefore, it is necessary for the BSs to synchronize themselves more frequently to be able to deliver those services.    

Figure \ref{fig:bpmesh} shows the RMSE of offset and skew estimation for scenario (b). As can be observed, the performance slightly deteriorates ($1-2$ ns for offset and $0.15-0.20$ ppm for skew) when compared to scenario (a). However, we note that the iterations of BRF are significantly faster than the iterations of BP. In particular, BP only begins when the nodes have already conducted $K$ rounds of time-stamp exchange (in order to form the matrices $\aij$ and $\aji$) and, even then, it still needs $4$ iterations (or $n$ iterations if there are $n$ nodes between a BS and MN) to perform synchronization. In contrast, BRF updates the estimation after each round of time-stamp exchange, thereby maintaining the relative clock offsets and skew low. In other words, since the BRF is faster (directly applied after each round of time-stamp exchange) and runs independently (does not need any other information from the other parts of the network as BP does), it is able to conduct more iterations, thereby continuously fulfilling the requirement of very low relative time error on a local level.

\section{Conclusions and Future Work}
We considered two Bayesian algorithms to estimate the clock offset and skew in communication networks. One is based on Belief Propagation and able to achieve reasonably accurate network-wide synchronization at the cost of a high number of time-stamp exchanges and message passing iterations, while the other is designed with the aid of Bayesian Recursive Filtering and capable of delivering superb performance in pairwise synchronization. Moreover, we employed both algorithms to construct a hybrid Bayesian approach to maintain a high sync accuracy on a global level while fulfilling the relative time error requirement at a local level. Simulation results show that the proposed hybrid network can achieve high precision and frequent offset and skew synchronization at the cost of only a slight deterioration in performance.

Furthermore, it is worth mentioning that precise time synchronization provides the basis for accurate localization. Therefore, our future work aims at designing localization algorithms based on the sync algorithms presented in this work. In particular, we will continue exploiting the benefits of hybrid approach to jointly synchronize and localize the nodes.
\bibliography{synch_paper_2}
\bibliographystyle{IEEEtran}
\end{document}

%% file: Macros.tex
\newcommand*{\argmax}{\ensuremath{\mathop{\mathrm{arg\,max}}}}
\newcommand{\tijt}[1]{\tilde{T}_{ij}^{#1}}
\newcommand{\rijt}[1]{\tilde{R}_{ij}^{#1}}
\newcommand{\tij}[1]{T_{ij}^{#1}}
\newcommand{\rij}[1]{R_{ij}^{#1}}
\newcommand{\muest}{\boldsymbol{\mu}_{\text{est}}}
\newcommand{\mucorr}{\boldsymbol{\mu}_{\text{update}}}
\newcommand{\mupred}{\boldsymbol{\mu}_{\text{pred}}}
\newcommand{\coest}{\mathbf{Q}_{\text{est}}}
\newcommand{\cocorr}{\mathbf{Q}_{\text{update}}}
\newcommand{\copred}{\mathbf{Q}_{\text{pred}}}
\newcommand{\ttat}[1]{\tilde{\boldsymbol{\vartheta}}_i^{#1}}
\newcommand{\tgam}{\frac{1}{\tilde{\gamma}_i}}
\newcommand{\tg}{\tilde{\gamma}_i}
\newcommand{\ttet}{\tilde{\theta}_i}
\newcommand{\uni}[1]{\mathbf{I}_{#1}}
\newcommand{\muvec}[3]{\boldsymbol{\mu}_{#1\rightarrow #2}^{(#3)}}
\newcommand{\sig}[3]{\boldsymbol{\Sigma}_{#1\rightarrow #2}^{(#3)}}
\newcommand{\omg}[3]{\boldsymbol{\Omega}_{#1\rightarrow #2}^{(#3)}}
\newcommand{\belcov}[2]{\mathbf{P}_{#1}^{(#2)}}
\newcommand{\belmu}[2]{\boldsymbol{\nu}_{#1}^{(#2)}}
\newcommand{\sigsel}[2]{\boldsymbol{\Sigma}_{#1}^{#2}}
\newcommand{\tetvec}[1]{\boldsymbol{\vartheta}_{#1}}
\newcommand{\tetvect}[1]{\tilde{\boldsymbol{\vartheta}}_{#1}}
\newcommand{\aij}{\mathbf{A}_{ij}}
\newcommand{\aji}{\mathbf{A}_{ji}}
\newcommand{\zji}{\mathbf{z}_{ij}}
\newcommand{\Cij}{\mathbf{C}_{ij}}
\newcommand{\ccij}{\mathbf{c}_{ij}}

\newcommand{\mij}[3]{\delta_{#1\rightarrow #2}^{(#3)}}
\newcommand{\pij}{p_{ij}}

\newcommand{\cntn}[2]{p(\ccij^{#1}|\tetvec{i}^{#2})}
\newcommand{\cntnt}[2]{p(\ccij^{#1}|\tetvect{i}^{#2})}

\newcommand{\tnt}[2]{p(\tetvect{i}^{#1}|\tetvect{i}^{#2})}
\newcommand{\tntt}[2]{p(\tetvect{i}^{#1}|\tetvect{i}^{#2})}

\newcommand{\tta}[1]{\boldsymbol{\vartheta}_i^{#1}}